\documentclass[aps,prd,10pt, twocolumn,superscriptaddress,nofootinbib,amsmath,amssymb,floatfix]{revtex4-2}
\pdfoutput=1

\usepackage{graphicx}
\usepackage{bm}
\usepackage{dcolumn}
\usepackage[dvipsnames]{xcolor}
\usepackage[normalem]{ulem}
\usepackage[makeroom]{cancel}
\usepackage[many]{tcolorbox}
\usepackage{diagbox}
\usepackage{float}
\usepackage{gensymb}
\usepackage{hyperref}

\newcommand{\diff}{\mathrm{d}}
\hypersetup{colorlinks=true, linkcolor=blue, citecolor=cyan}

\begin{document}

\title{Timing of echoes in electromagnetic radiation from hot spots\\
orbiting a Kerr black hole impostor}

\author{Sudipta Hensh}
\email{sudiptahensh2009@gmail.com}
\author{Jan Schee}
\email{jan.schee@physics.slu.cz}
\author{Zden\v{e}k Stuchl\'{i}k}
\email{zdenek.stuchlik@physics.slu.cz}
\affiliation{Research Centre for Theoretical Physics and Astrophysics,
Institute of Physics, Silesian University in Opava,
Bezru\v{c}ovo n\'{a}m\v{e}st\'{i} 13, CZ-74601 Opava, Czech Republic}

\date{\today}

\begin{abstract}
We investigate the light curve of an isolated bright spot in a Keplerian orbit around a Kerr black hole candidate to probe possible deviations from the standard event-horizon picture. In particular, we explore the presence of a reflective surface, namely a ``mirror" near the horizon, and its observable consequences. 
We present estimates on the time delays of the echoes of electromagnetic radiation from the mirror or some other reflecting area related to black hole mimickers.
Our results show that photon reflections from such a surface imprint distinct late-time features in the light curve, producing an electromagnetic analogue of gravitational wave echoes. For lower values of the spin, the reflected signal appears within the same orbital period and leads to a noticeable enhancement of the flux. In contrast, for higher spins, the effect emerges only after several orbits, giving rise to a characteristic echo-wave train. These features provide a potential observational signature of horizon-scale modifications in compact objects.
\end{abstract}
\maketitle

\section{Introduction}

Over the past decade, technological advancements in probing astrophysical phenomena have transformed our understanding of compact objects. Gravitational wave detections by the LIGO–Virgo–KAGRA collaboration established the existence of astrophysical black holes~(BHs)~\cite{2016}. Shortly thereafter, the Event Horizon Telescope (EHT) observed the first image of the supermassive BH in the galaxy M87, providing direct evidence of a horizon-scale, ultra-compact object~\cite{EventHorizonTelescope:2019dse,EventHorizonTelescope:2019ths}. 
In addition, the polarized image of the M87 BH~\cite{EventHorizonTelescope:2021bee,EventHorizonTelescope:2021srq} has offered new insights into the magnetic field structure near the event horizon. 
While these achievements marked major milestones, they have not yet fully resolved the distinction between classical BHs and horizonless compact object mimickers.
Looking ahead, next-generation ground-based detectors such as the Einstein Telescope and Cosmic Explorer~\cite{Punturo:2010zz,Reitze:2019iox}, the space-based LISA mission~\cite{2017arXiv170200786A}, together with upgrades of the EHT~\cite{2019BAAS...51g.256D}, are expected to deliver more stringent tests of strong-field gravity and may help to reveal the true nature of BHs.

Vacuum fluctuations produce particle-antiparticle entangled pairs, and they annihilate very rapidly because of their short lifespan. 
When this pair production occurs in the close vicinity of the event horizon, then one of them falls into the BH, and the other one escapes to infinity as Hawking radiation~\cite{Hawking:1975vcx}. 
Information loss in Hawking radiation is still an active area of research. A proposal to avoid information loss in Hawking radiation is to give up on Einstein's principle of equivalence by considering that the in-falling observer would observe a pile of high-energy particles in the close vicinity of the horizon; namely, firewall~(see~\cite{Almheiri:2012rt} for details).
Such phenomenological models give us a motivation to study horizonless BH mimickers endowed with a reflecting mirror.

The existence of such a reflective surface~(we refer to it as ``mirror'' hereafter) would basically appear to be reflected in the strong field signatures that we can observe due to astrophysical events involving BHs. 
This can be done mainly in two possible ways: in the gravitational wave emission or in the observed electromagnetic emissions. 
In this article, we intend to investigate the latter scenario. We are interested in observing signatures of such a mirror in the electromagnetic spectrum of a radiating hot spot. 
% Any intrinsic variability in the local emission from the innermost region of the accretion disk could influence the observed X-ray variability~\cite{1991A&A...245..454A}.
The observational electromagnetic signatures of a corotating spot in an accretion disk were studied in~\cite{1992A&A...257..594B} and the radiation from hot spots in the extreme Reissner-Nordstrom BH was investigated in~\cite{Stuchlik:1992phc}. 

Detection of hot spot orbiting in a stable circular orbit around our galactic center BH SgrA* by the Gravity Collaboration~\cite{refId0} has introduced a lot of curiosity in the scientific community to model the hot spot. There are many efforts available in the literature to study the hot spot. 
While we can not do justice to mention all the efforts, here we refer to some efforts such as~\cite{10.1093/mnras/269.2.283,2004AIPC..714...40S,2000MNRAS.315....1R,1996ApJ...470..743K,1995NYASA.759..550Z,2006AN....327..957P,2003PASJ...55.1121F,1994ApJ...425...63B,Bao-Stu:1992:ApJ}. 
A work dedicated to distinguishing between BHs and worm holes using a hot spot can be found in~\cite{Li:2014coa}. 
While the model of hot spot from plasma microphysics is still an active area of research, our goal in this article is not to understand the underlying microphysics that governs the formation of a hot spot, but to focus on observational imprints of the mirror on the electromagnetic spectrum.

It is worth emphasizing that the framework developed here is not limited to Kerr BHs with a putative reflective mirror; we refer to the system as ``Kerr mirror'' hereafter. Our investigation is equally applicable to a broader class of compact objects that mimic BH properties but possess a hard surface rather than an event horizon, including gravastars~\cite{Mazur:2001fv,Visser:2003ge}, boson stars~\cite{Liebling:2012fv}, and other horizonless exotic compact objects~\cite{Cardoso:2019rvt,Stu-Vrb:2026:PhRvD,Stu-Vrb-Tur:2026:EPJC}. 
In these scenarios, the presence of a physical surface naturally allows for the reflection of radiation, leading to signatures in the observed light curves analogous to the echo-like features we describe.
A similar consideration applies to a neutron star~(NS) with a partially ionized
atmosphere, which can reflect part of the incoming radiation, as shown in
\cite{2012A&A...546A.121P}. It is instructive to compare an NS with a compact
object of the same mass $M$ whose reflecting surface lies just outside the
gravitational radius $r_{+} = 2M$. 
In both cases, the illuminating source is placed at the same distance $d$ from the centre of the object, and both surfaces are assumed to have the same reflectivity, so that any difference in the reflected signal is purely geometrical.
The comparison is most naturally phrased in terms of the compactness
$\mathcal{C} \equiv R_{\rm NS}/M$, which is dimensionless and therefore
independent of the mass scale. For realistic equations of state
$\mathcal{C}$ lies in the range $\mathcal{C} \simeq 4$--$6$, to be compared
with $r_{+}/M = 2$ for the BH case.
A sphere of radius $R$ viewed from a distance $d$ subtends the solid angle
% = 2\pi \left( 1 - \sqrt{1 - R^{2}/d^{2}} \right)
$\Omega \simeq \pi R^{2}/d^{2}$
for $d \gg R$, i.e., the projected area of the reflecting surface divided by the
square of the source distance. Since $d$ is, by assumption, the same for both
objects, the intercepted fraction of the emitted radiation scales as $R^{2}$,
giving $\Omega_{\rm NS}/\Omega_{\rm BH} \simeq (R_{\rm NS}/r_{+})^{2}
= (\mathcal{C}/2)^{2} \approx 4$--$9$, independently of the mass. 
The NS therefore intercepts several times more of the incident radiation, and for
equal reflectivity, the reflected component --- and consequently the echo signal
--- is expected to be correspondingly stronger.
The exterior geometry of an NS, however, differs from the Kerr
solution, and modelling the reflection from a partially ionized atmosphere
would require assumptions specific to the stellar structure and atmospheric
composition. 
Since the present work is concerned with the properties of BH mimickers, for which a wide range of physically motivated models is available in the literature, we restrict the following discussion to the BH case and leave the NS scenario for future study.

It is worth noting that the presence of a mirror around the Kerr BH does not affect the spacetime under consideration.
Thus, while the Kerr spacetime serves as the primary background in this work, the phenomenology we uncover is more general and provides a potential observational window into distinguishing BHs from their mimickers.

In Sect.~\ref{eom}, we describe the governing equations of motion that uniquely determine the path traversed by the photon, that is, a geodesic. 
Next in Sect.~\ref{ray-tracing}, we describe our mirror model considering how this investigation is done and ray-tracing in the presence of the mirror. 
The construction of the light curve is described, and a comparison plot of the light curve is demonstrated in Sec~\ref{light_curve}. The optical effect of a mirror is discussed in Sect.~\ref{echo_sec}. In Sect.~\ref{discussion}, a discussion of the whole analysis is presented. 
Throughout the paper, unless otherwise mentioned, we consider $(-,+,+,+)$ signatures and geometrical system of units, i.e., $G=c=1$.

\section{Equations of motion}\label{eom}

The spacetime interval of Kerr geometry in the Boyer-Lindquist coordinates~$(t, r, \theta, \phi$) reads~ \cite{Bardeen1972}
\begin{eqnarray}
    \diff s^2&=&-\left(1-\frac{2Mr}{\Sigma}\right)\diff t^2 +\frac{\Sigma}{\Delta}\diff r^2 + \Sigma\,\diff\theta^2 \nonumber \\
    && +\frac{A}{\Sigma}\sin^2\theta\diff\phi^2-\frac{4a\,r}{\Sigma}\sin\theta\diff t\diff\phi
\end{eqnarray}
where
\begin{eqnarray}
	\Delta&=&r^2-2M\,r+a^2,\\
	\Sigma&=&r^2+a^2\cos^2\theta,\\
	A&=&(r^2+a^2)^2-a^2\Delta\sin^2\theta \ ,
\end{eqnarray}
where $a$ is the spin of the BH. For numerical advantages, we use new radial and new latitudinal coordinates defined by $u\equiv 1/r$ and $m\equiv\cos\theta$. Following the Hamilton-Jacobi separation of variables procedure, one can easily get the Carter equations of motion \cite{Bardeen1972,Carter1968}. The equations of motion of photon with 4-momentum $k^{\mu}$ are given by
\begin{eqnarray}
    \Sigma k^t&=& a Q(l,m,a) + \frac{(1+u^2a^2) P(u,l,a)}{u^2 \Delta} \label{carter_t} \ ,\\
	\Sigma k^u&=& \pm \sqrt{U(u,l,q,a)} \label{carter_u} \ ,\\
	\Sigma k^m&=& \pm \sqrt{M(m,l,q,a)} \label{carter_m}  \ ,\\
	\Sigma k^{\phi}&=& \frac{Q(l,m,a)}{1-m^2} + \frac{a P(u,l,a)}{\Delta} \label{carter_phi} \ ,
\end{eqnarray}
with $l\equiv -k_\phi/k_t$ and $q\equiv \mathcal{C}-l^2$ ($\mathcal{C}$ is Carter separation constant
) being constants of motion, and we define
\begin{eqnarray}
    Q(l,m,a)&\equiv& l-a(1-m^2)  \ ,\\
    P(u,l,a)&\equiv& \frac{(1+u^2 a^2)}{u^2} - la \ ,\\
	U(u,l,q,a)&\equiv&1+(a^2-l^2-q)u^2+2[(a-l)^2+q]u^3 \label{capU}\nonumber \\
    &&-a^2\,q\,u^4 \ ,\\
	M(m,l,q,a)&\equiv&q+(a^2-l^2-q)m^2-a^2 m^4 \ .
\end{eqnarray}
 Introducing a new geodesic's parameter $\tilde{\lambda}$ by rescaling as
\begin{equation}
	\frac{\diff}{\diff\tilde{\lambda}}\equiv\Sigma\frac{\diff}{\diff\lambda} \ ,
\end{equation}
where $\lambda$ parametrizes the geodesic as proper time for timelike curves, but for null geodesics it serves as an affine parameter, measuring progress along the light ray without representing elapsed time on any physical clock.
Since $\Sigma$ depends on position, $\tilde{\lambda}$ is no longer an affine parameter but a general coordinate parameter, we introduce $\tilde{k}^\mu = dx^{\mu}/d\tilde{\lambda}$.

With this new parameter equations~(\ref{carter_t})-(\ref{carter_phi}) take the form
\begin{eqnarray}
\tilde{k}^t&=& a Q(l,m,a) + \frac{(1 + u^2 a^2) P(u,l,a)}{u^2 \Delta} \label{carter_t2} \ , \\
\tilde{k}^u&=& \pm \sqrt{U(u,l,q,a)}\label{carter_u2} \ , \\
\tilde{k}^m&=& \pm \sqrt{M(m,l,q,a)}\label{carter_m2} \ , \\
\tilde{k}^{\phi}&=& \frac{Q(l,m,a)}{1-m^2} + \frac{a P(u,l,a)}{\Delta} \label{carter_phi2} \ .
\end{eqnarray}

In order to eliminate the square roots in Eqs.~\eqref{carter_u2} and \eqref{carter_m2}, we reformulate them as second-order differential equations, leading to a set of equations together with Eqs.~\eqref{carter_t2}, \eqref{carter_phi2} employed in our numerical code as follows,
% \begin{eqnarray}
% 	\frac{\diff k^u}{\diff\lambda'}&=&\frac{1}{2}\frac{\diff U}{\diff u} \ ,\label{kueq}\\
% 	\frac{\diff k^m}{\diff\lambda'}&=&\frac{1}{2}\frac{\diff M}{\diff m} \ .\label{kmeq}
% \end{eqnarray}
\begin{eqnarray}
    % \frac{dk^u}{d\lambda'}&=&\frac{1}{2}\frac{dU}{du},\\
    % \frac{dk^m}{d\lambda'}&=&\frac{1}{2}\frac{dM}{dm},\\
    \frac{du}{d\tilde{\lambda}}&=&\tilde{k}^u \ , \label{ku}\\
    \frac{dm}{d\tilde{\lambda}}&=&\tilde{k}^m \ ,\label{km}\\
    \frac{\diff \tilde{k}^u}{\diff\tilde{\lambda}}&=&\frac{1}{2}\frac{\diff U}{\diff u} \ ,\label{kueq}\\
	\frac{\diff \tilde{k}^m}{\diff\tilde{\lambda}}&=&\frac{1}{2}\frac{\diff M}{\diff m} \ .\label{kmeq}
    % \frac{d\phi}{d\lambda'}&=& \frac{Q(l,m,a)}{1-m^2} + \frac{a P(u,l,a)}{\Delta} \ ,\\
    % \frac{dt}{d\lambda'}&=& a Q(l,m,a) + \frac{(u^2+a^2) P(u,l,a)}{u^2 \Delta}.
\end{eqnarray}
% The corresponding raytracing procedure is explained in the following section.

%\pagebreak

\section{Ray-tracing and reflection off the mirror} \label{ray-tracing}

We consider the mirror as a reflecting surface situated at $r_f=r_h+\epsilon$, i.e., just above the event horizon~($r_h$). 
To keep the problem simple and highlight qualitatively with maximal effect of the reflection from the mirror, we consider the mirror as a perfect reflector, and the source of the radiation is only from the hot spot. In addition, we also assume that the hot spot is geometrically thin and optically thick.
In our model, we also consider that the hot spot is confined to the equatorial plane, maintaining a circular orbit and having a finite radial extension  $dr$.

Here we are using the so-called backward ray-tracing method. Using this technique, one can trace the rays from the detector location to the source of emission.
At first, we construct a two-dimensional photographic plate, and the coordinates $(\alpha,\beta)$ determine the location of the pixel on the photographic plate.

Each geodesic can be determined by the two coordinates $(\alpha,\beta)$, tightly connected with impact parameters $l$ and $q$. The relation between the photographic plate coordinate and the constants of motion reads~\cite{1973ApJ...183..237C}

\begin{eqnarray}
	l&=&\alpha\sin\theta_o,\label{alpha1}\\
	q&=&\beta^2+\cos^2\theta_o\left(\alpha^2-a^2\right),\label{beta1}
\end{eqnarray}
where $\theta_o$ is the inclination angle of the observer.
In order to determine the origin of the observed photon, we construct its geodesic by taking the following steps:

\begin{itemize}
    
\item First, we estimate the range of the photographic plate coordinates by taking a projection of the whole BH-hot spot system onto the observer's two-dimensional sky.

\item Then we divide the photographic plate $(\alpha-\beta)$ into finite regions (pixels) having dimensions $\Delta\alpha\times\Delta\beta$ where $\Delta\alpha\equiv(\alpha_{max}-\alpha_{min})/N_\alpha$ and $\Delta\beta\equiv(\beta_{max}-\beta_{min})/N_\beta$, where $N_\alpha$ and $N_\beta$ are integers. For a given pixel $\alpha\in[\alpha_{min},\alpha_{max}]$ and $\beta\in[\beta_{min},\beta_{max}]$ determine $l$ and $q$ from equations (\ref{alpha1}) and (\ref{beta1}).

\item Next, we calculate the turning point $u_t$ i.e. the smallest positive root of $U(u, l, q, a) = 0$, with $U$ given by Eq.~\eqref{capU}. 
Those turning points that occur between the radius of the outer edge of the hot spot, $r_{\mathrm{max}}$, and the radius of the event horizon, $r_h$, are of our interest. This condition can be expressed as $u_t> u_{\mathrm{max}} \equiv 1/r_{max}$ and $u_t< u_h \equiv 1/r_h \equiv (1+\sqrt{1-a^2})^{-1}$. 

\item In order to be computationally efficient, assuming the hot spot spans between $r_1$ and $r_{\mathrm{max}}$, we determine the maximal meaningful value of the  parameter $\tilde{\lambda}_{max}$ as

	\begin{equation}
		\tilde{\lambda}_{max}=\left\{
						\begin{array}{cc}
							\int_{u_o}^{u_h}{\frac{\diff u}{\sqrt{U(u;l,q,a)}}} & \textrm{ for } n_u=0 \ ,\\
							\int_{u_o}^{u_t}{\frac{\diff u}{\sqrt{U(u;l,q,a)}}} +\int_{u_{\mathrm{max}}}^{u_t}{\frac{\diff u}{\sqrt{U(u;l,q,a)}}} & \textrm{ for } n_u=1 \ ,
						\end{array}
		\right.
	\end{equation}

where $u_o=1/r_o$ is the distance at which the observer is located, and $n_u$ indicates the number of existing turning points along the geodesics (in the case of null geodesics, we have two possibilities: $n_u=0$, indicating that there is no turning point, and $n_u=1$, indicating that there is one turning point). 

\item To get the geodesic we solve equations \eqref{carter_t2}, \eqref{carter_phi2},
%\eqref{ku}, \eqref{km},
\eqref{kueq}, \eqref{kmeq} with the initial conditions at $\tilde{\lambda}=0$ as

\begin{eqnarray}
	t(0)&=& 0 \ , \\
	u(0)&=&u_o \ ,\\
	m(0)&=&m_o \ ,\\
	\phi(0)&=& 0 \ , \\
	\tilde{k}^u(0)&=&\sqrt{U(u_o;l,q,a)} \ ,\\
	\tilde{k}^m(0)&=&\mathrm{Sign}(\beta)\sqrt{M(m_o;l,q,a)} \ ,
\end{eqnarray}

where $m_o=\cos \theta_o$ is the inclination angle of the observer.

\item Now the procedure splits into two cases: one with no turning point and the other with a turning point.

\end{itemize}

\begin{figure*}[ht]
    \centering
    \begin{tabular}{cc}
    \includegraphics[width=0.95\linewidth]{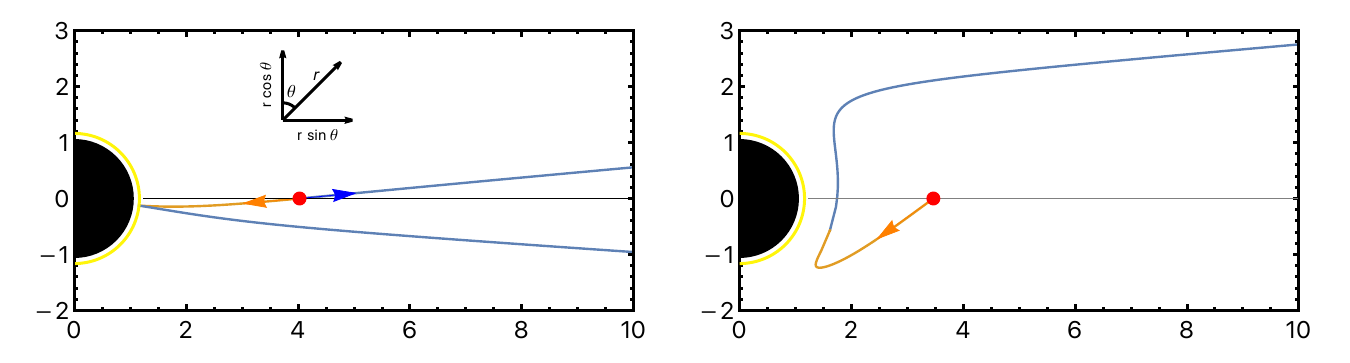}
    \end{tabular}
    \caption{The illustration of radiation propagation from a hot spot (the red dot ) in the field of the Kerr mirror,  represented by a yellow half circle surrounding the Kerr event horizon (a black half disk). On the left panel, null geodesics with no turning points are plotted. There are two photons with the same impact parameters, $l$ and $q$, corresponding to different $\alpha$ and $\beta$ coordinates in the image plane, one is emitted outwards (blue) and the second inwards (orange). The orange geodesics reflect off the Kerr mirror and travel back to infinity.  On the right panel, the null geodesics having a turning point are plotted. It is divided into two segments separated by the corresponding radial turning point.  }
    \label{fig:reflection}
\end{figure*}
	
If there is no turning point, i.e., $n_u=0$, then we integrate equations of motion down to the horizon, where the solution reads

\begin{eqnarray}
    t(\tilde{\lambda}_{max})&=& t_h \ , \\
	u(\tilde{\lambda}_{max})&=&u_h \ ,\\
	m(\tilde{\lambda}_{max})&=&m_h \ ,\\
	\phi(\tilde{\lambda}_{max})&=& \phi_h \ , \\
	\tilde{k}^u(\tilde{\lambda}_{max})&=&\tilde{k}^u_h \ ,\\
	\tilde{k}^m(\tilde{\lambda}_{max})&=&\tilde{k}^m_h \ .
\end{eqnarray}
	
We apply to this propagation vector a perfect (energy and angular momentum are conserved) reflection by converting the radial component of the momentum from global coordinate~(GC) to the locally non-rotating frame~(LNRF), applying the reflection, and then switching back to the GC frame as given below (the transformation equations were first introduced in \cite{Bardeen1972})

\begin{eqnarray}
    \tilde{k}^u_h \xrightarrow{\textrm{LNRF}} \tilde{k}'^u_h \xrightarrow{\textrm{Reflection}} - \tilde{k}'^u_h \xrightarrow{\textrm{GC}} \tilde{k}''^u_h \ ,
\end{eqnarray}
and we assume that the ``mirror" is rotating as LNRF, which is thus the frame of reflection.
	
So the new initial conditions read
\begin{eqnarray}
	t(0)&=& t_h \ , \\
	u(0)&=&u_h,\\
	m(0)&=&m_h,\\
	\phi(0)&=& \phi_h \ , \\
	\tilde{k}^u(0)&=& \tilde{k}''^u_h \ , \\
	\tilde{k}^m(0)&=&\tilde{k}^m_h.
\end{eqnarray}

We determine the new maximal value of the $\tilde{\lambda}$ parameter
\begin{equation}
\tilde{\lambda}_{max}=\int_{u_h}^{u_{\textrm{max}}}\frac{\diff u}{\sqrt{U(u;l,q,a)}} \ .
\end{equation}	 
Then we integrate the system of differential equations  \eqref{carter_t2}, \eqref{carter_phi2}, \eqref{kueq}, \eqref{kmeq}, and check the intersection between the rays and the hot spot.

If there is a turning point $n_u=1$ in this case, the geodesic is not divided into segments. We integrate system of differential equations  \eqref{carter_t2}, \eqref{carter_phi2}, \eqref{kueq}, \eqref{kmeq} down to $\tilde{\lambda}=\tilde{\lambda}_{max}$ and check for intersection with the hot spot. 
The reflection off the Kerr mirror and an example of null geodesics with a turning point are illustrated in Fig. \ref{fig:reflection}.

\section{Light Curve} \label{light_curve}
We initiate our simulations by performing backward ray tracing of null geodesics from the observer’s screen into the Kerr spacetime. Along each geodesic, we determine its intersections with the equatorial plane, which represents the orbital plane of the hot spot. At every such intersection, we identify the corresponding emission radius $r_e$. In addition, we compute the associated frequency shift factor $g=(k_{\mu}v^\mu)_o / (k_\mu v^\mu)_e$ (``o" refers to observer, ``e"-refers to source, $v^\mu$ is 4-velocity), and it reads  \cite{Fan-etal:1997PASJ,Schee-Stu:2009GRG}
\begin{eqnarray}
	g =\frac{\left[1-\frac{2}{r_e}(1-a\,\Omega_e)^2-(r_e^2+a^2)\Omega_e^2\right]^{1/2}}{
    \left(1-l\Omega_e\right)} \label{freq_shift_ok}  ,
\end{eqnarray}
where $\Omega_e$ is the orbital angular velocity of the emitter in a Keplerian orbit, which can be written as
\begin{eqnarray}
    \Omega_e = \frac{1}{r_e^{3/2}+a} \label{eq:Omega}.
\end{eqnarray}

This frequency shift factor encapsulates both gravitational and kinematic contributions.

In our simulations, the hot spot is assumed to orbit the central compact object along a circular geodesic. The local emissivity profile of the hot spot is modeled through the intensity function~\cite{Schnittman:2003tm}
\begin{equation}
    I_e = \epsilon_s e^{-d^2/2\sigma^2} \ , \label{eq:I}
\end{equation}
where $d$ denotes the local spatial distance from the center of the hot spot to a given emission point, $\epsilon_s$ represents the central (peak) intensity, and $\sigma$ is the standard deviation characterizing the width of the Gaussian distribution. This choice of profile corresponds to a geometrically thin, optically thick hot spot with a Gaussian falloff in brightness, ensuring that the emission is concentrated near the hot spot center while smoothly decreasing toward its edges. Such a parametrization provides a simple yet physically motivated description of localized emission regions and has been widely employed in hot spot models in the literature. 

When a photon intersects the equatorial plane at coordinates $(r_e,\phi_e)$, the spatial separation from the hot spot center is required to evaluate the intensity profile given in Eq.~\eqref{eq:I}. 
Neglecting relativistic corrections to proper length, we approximate this distance $d$ using the cosine law in the equatorial plane, obtaining
\begin{equation}
    d^2 = r_s^2 + r_e^2 - 2 r_s r_e \cos(\phi_s - \phi_e), \label{eq:d}
\end{equation}
where $r_s$ denotes the orbital radius of the hot spot center and $\phi_s$ its azimuthal coordinate. The quantity $d$ thus measures the Euclidean distance between the photon’s emission point and the hot spot center projected onto the equatorial plane. This approximation, while neglecting curved spacetime corrections to proper length, provides a simple and effective means to characterize the hot spot’s spatial emissivity distribution within our model.

\begin{figure*}[ht]
    \centering
    \includegraphics[width=0.95\linewidth]{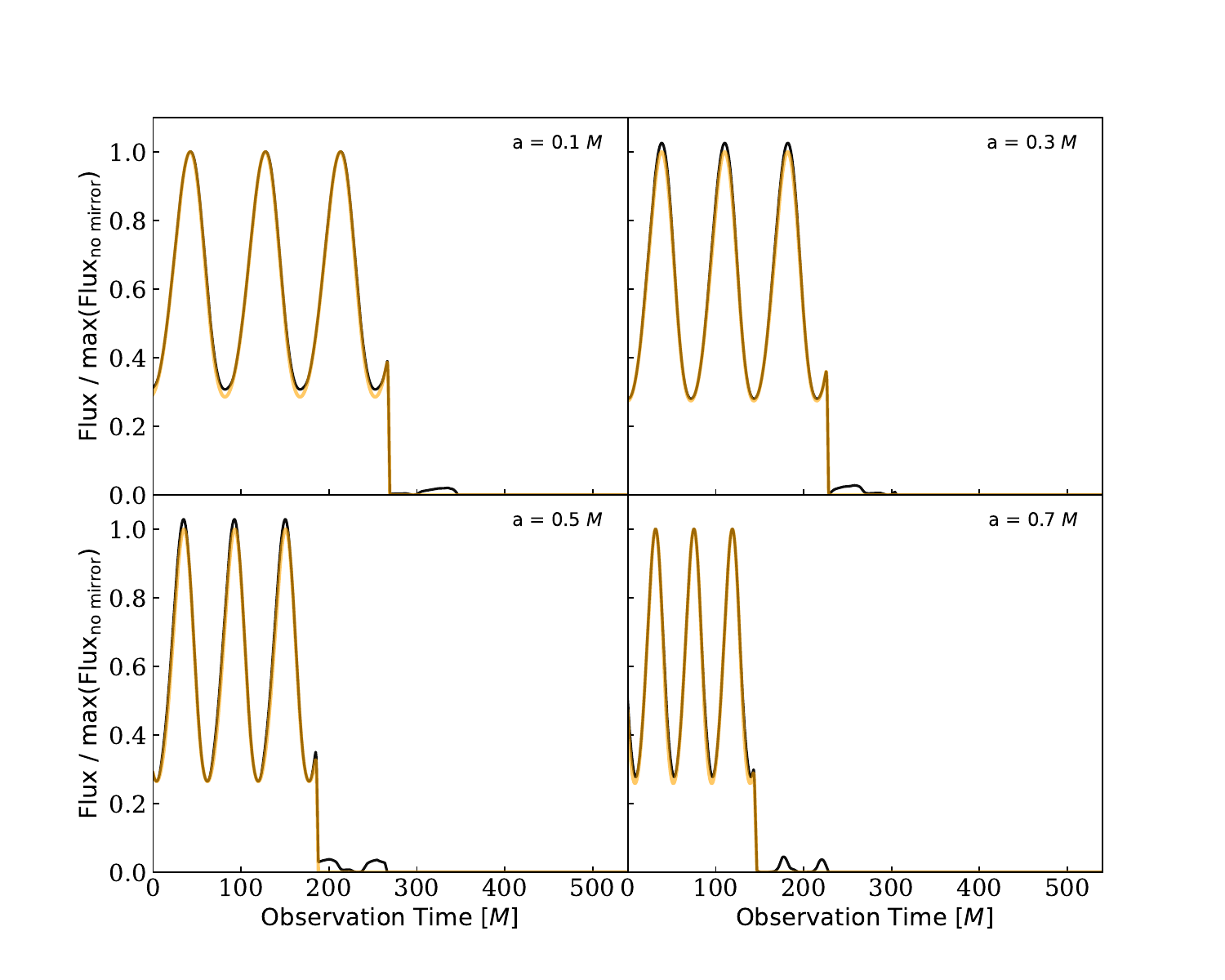}

    \caption{Comparison of light curves for a hot spot located at $r=r_{_{\mathrm{ISCO}}}$ for Kerr and mirror scenarios considering different BH spins, observer's inclination angle $\theta=30\degree$, and mirror depth $\epsilon=10^{-6}M$. The yellow curve represents the standard Kerr spacetime, while the black curve shows the modified emission profile in the mirror scenario.}
    \label{fig:spineffect}
\end{figure*}

\begin{figure*}
    \centering
    \includegraphics[width=0.95\linewidth]{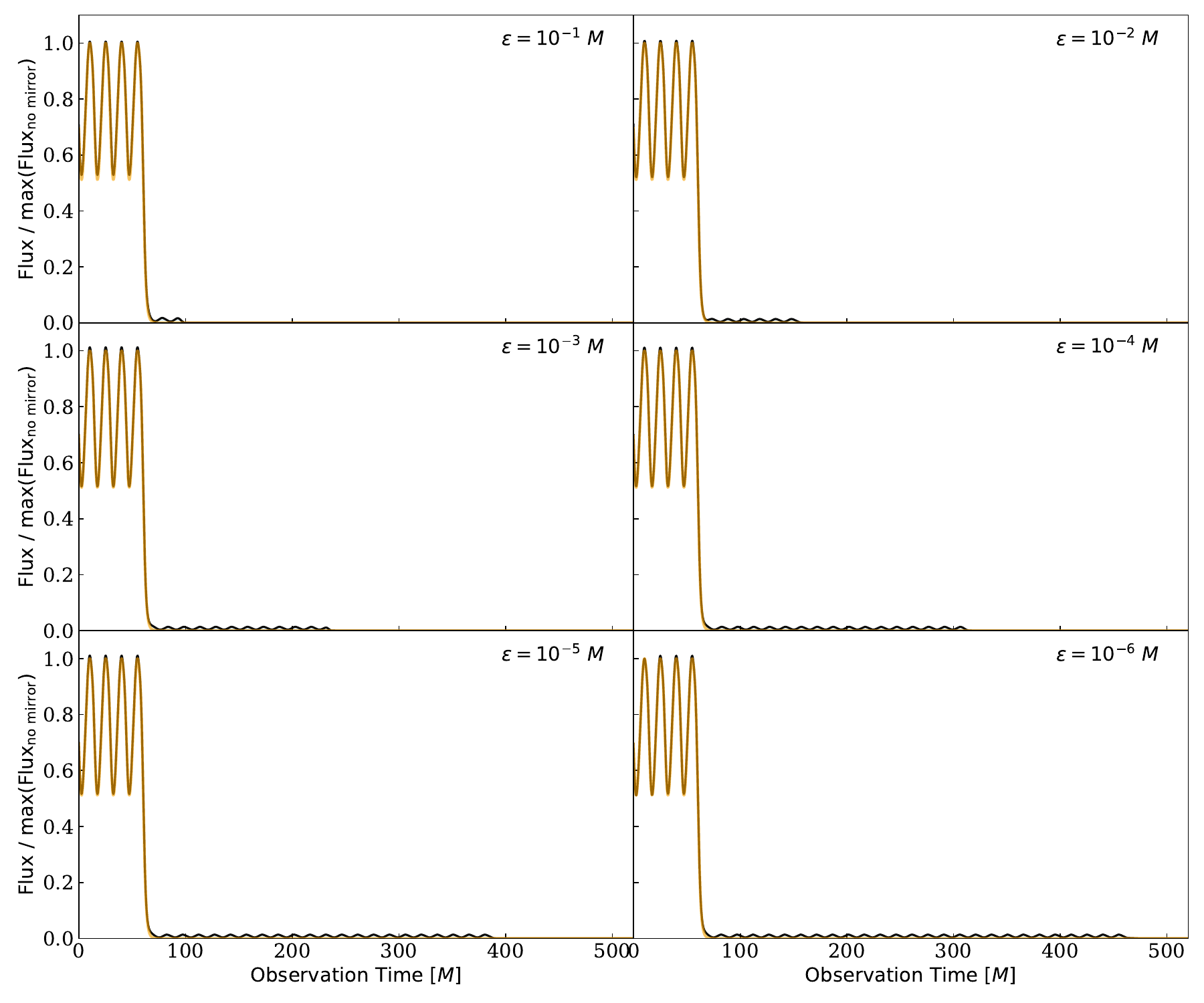} 			  
    
    \caption{Comparison of light curves for hot spot located at $r=r_{_{\mathrm{ISCO}}}$ for Kerr and mirror scenarios considering different mirror depths $\epsilon$ for spin $a=0.998M$ with observer's inclination angle $\theta=30\degree$. 
    The yellow curve represents the standard Kerr spacetime, while the black curve shows the modified emission profile in the mirror scenario.}
    \label{fig:maxspindiffeps}
\end{figure*}

In order to construct the light curve, we collect the photons detected by a distant observer and sort them into time bins. The observation time is assumed to lie within the interval $t_o \in [t_{\min},t_{\max}]$, which we divide into $N$ equal bins of size  
\begin{equation}
    \Delta t \equiv \frac{t_{\max}-t_{\min}}{N}.
\end{equation}
For each time bin $[t_{o(i)},\, t_{o(i)}+\Delta t]$ with bin index $i$, we determine the contribution from all simulated photons ($j \in 0,1,\dots,N_\gamma$, where $N_\gamma$ is the total number of photons) emitted from the equatorial plane with radial coordinates lying in the range $[r_{\mathrm{ISCO}},\, r_{\max}]$, as the inner edge of the hot spot is located at the innermost stable circular orbit, namely, ISCO.
The procedure can be summarized as follows:  

\begin{itemize}
    \item For each photon, we define the emission time relative to the earliest photon (identified with a photon index value $m$) arrival as  
    \begin{equation}
        \Delta t_e = \Delta t_{e(j)} - \min\!\left(\Delta t_{e(m)}\right),
    \end{equation}
    and record its emission radius and azimuthal coordinate as  
    \begin{equation}
        r_e = r_{e(j)}, \qquad \phi_e = \phi_{e(j)}.
    \end{equation}
    The corresponding emission time is related to the observation time~($t_o$) as  
    \begin{equation}
        t_e = t_o + \Delta t_e \ .
    \end{equation}

    \item Using the emission coordinates $(r_e,\phi_e)$, we compute the spatial separation $d$ from the hot spot center according to Eq.~\eqref{eq:d}, and evaluate the corresponding local intensity $I_e$ using Eq.~\eqref{eq:I}.  

    \item We calculate the orbital angular frequency of the hot spot center, $\Omega(r_s)$, from Eq.~\eqref{eq:Omega}, and the frequency shift factor $g$ from Eq.~\eqref{freq_shift_ok}.  

    \item Finally, the observed flux contribution of each photon is accumulated in the corresponding time bin according to  
    \begin{equation}
        F_{o(i)} := F_{o(i)} + g^4 I_{e(j)} \ ,
    \end{equation}
    where  $:=$ represents the value assignment operation (the right-hand-side  is evaluated and copied into the expression on the left-hand-side of the $:=$  operation). The quantity $F_{o(i)}$ represents the $i$-th bin of observed flux, $I_{e(j)}$ is the source intensity corresponding to the $j$-th record in the table of all collected photons.

\end{itemize}
This procedure effectively maps the simulated photon trajectories and their associated redshifted intensities into an observed light curve. The factor $g^4$ accounts for (i) the energy shift of photons, (ii) the loss of the rate of photon numbers due to the time delay effect, and (iii) solid angle measurement relativistic corrections in curved spacetime, ensuring a consistent transformation from the emitter’s frame to the observer’s frame.

\section{The Echo}\label{echo_sec}

\begin{figure*}
    \centering
    \includegraphics[width=0.95\linewidth]{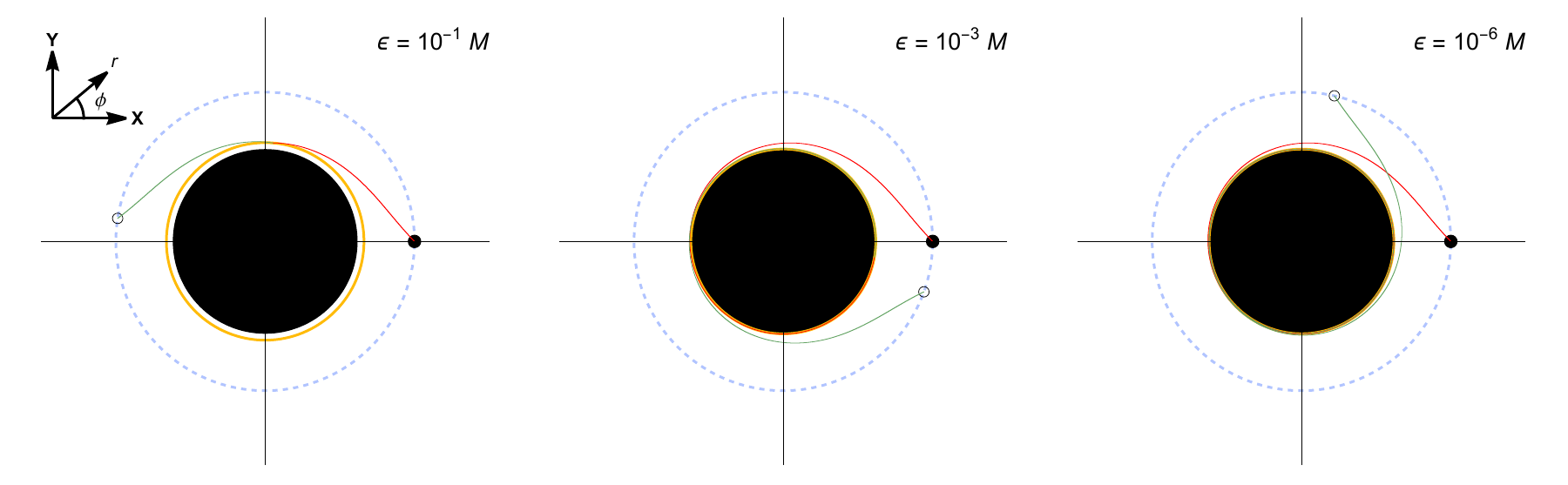}
    \includegraphics[width=0.95\linewidth]{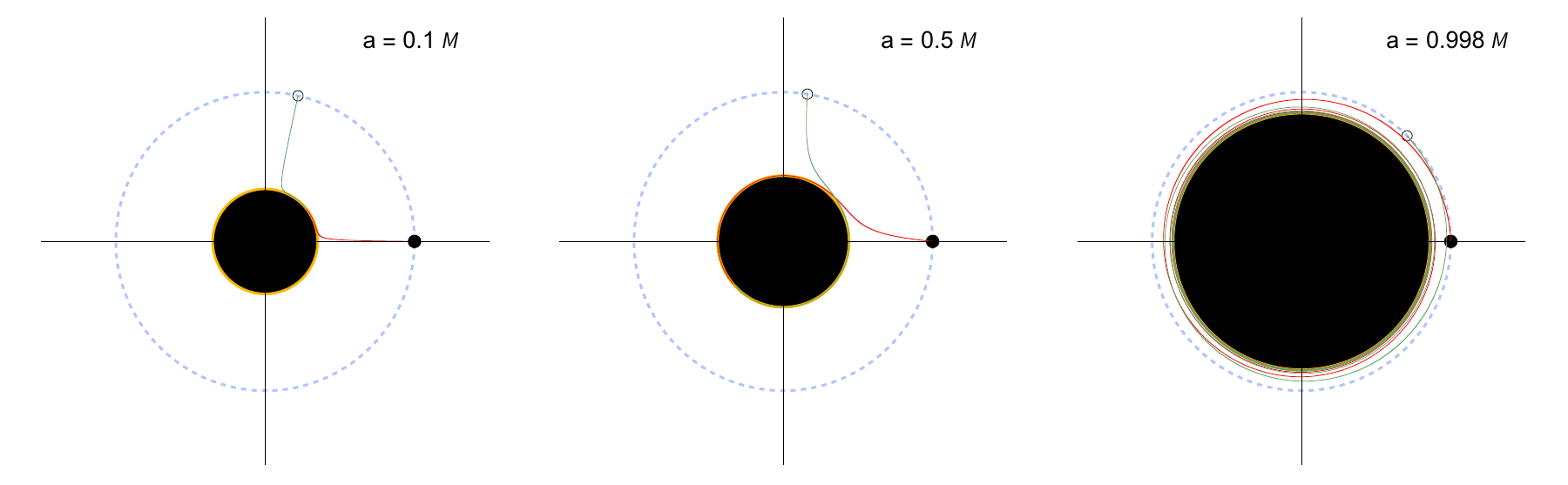}
   \caption{Trajectories of photons in the equatorial plane ($\theta=\pi/2$) traveling from the innermost stable circular orbit~(ISCO) to the mirror where they undergo reflection. Different scenarios are illustrated for a fixed value of the spin $a=0.9M$ with various $\epsilon$ ({\it upper panel}), and for a fixed value of $\epsilon=10^{-6}M$ with various spin values $a$ ({\it lower panel}). The black disk represents the BH, the orange circle marks the mirror location, and the light-blue dashed circle corresponds to the ISCO. Photons are emitted from the black dot (follow the red line), encounter the mirror, and are reflected and return to the hollow circles (following the green line), which mark their return points at the ISCO.}
   \label{fig:echoI}
\end{figure*}

\begin{figure}[ht]
    \centering
    \includegraphics[width=0.95\linewidth]{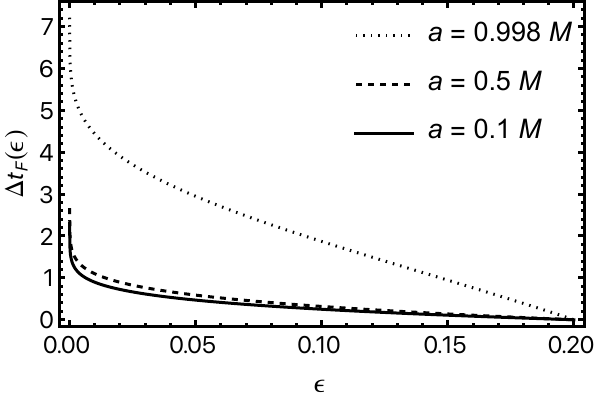}
    \caption{Echo delay for a photon to travel from the innermost stable circular orbit~(ISCO) to the mirror and back, plotted as a function of $\epsilon$ for different spins $a$.}
    \label{fig:echoII}
\end{figure}

\begin{figure*}[ht]
    \centering
    \includegraphics[width=0.98\linewidth]{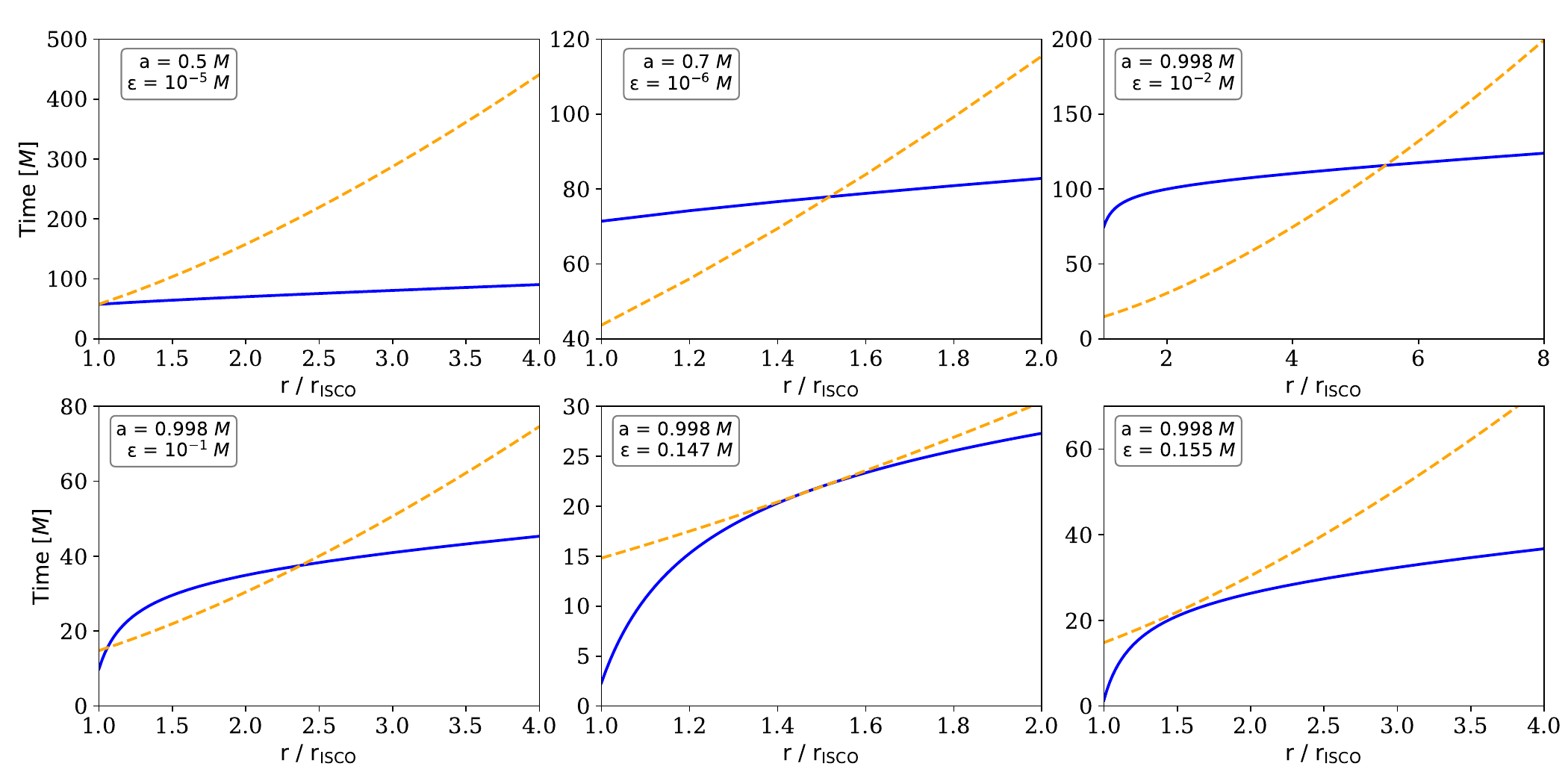}
    \caption{Echo delay for a radial photon to complete a round trip between the hot spot location and mirror (solid blue line) calculated from formula (\ref{eq:delay}), compared with the orbital period at each radius (dashed yellow line). Both quantities are plotted as functions of radius in units of the ISCO radius.}
    \label{fig:echoIII}
\end{figure*}

The presence of a reflective surface, a mirror, near the horizon leads to the appearance of an echo signal in the observed light curve. This echo manifests as a typically weak contribution that persists long after the primary hot spot emission has ceased~(Figs.~\ref{fig:spineffect}, \ref{fig:maxspindiffeps}). The duration and strength of the echo depend sensitively on the location of the mirror: the deeper (i.e., closer to the horizon) the reflecting surface is placed, the longer the associated delay in the reflected signal.
Let us discuss the issue of the observability of the echo.  Assume NuSTAR observatory with its instruments sensitivity $\Delta F\sim 10^{-14}\mathrm{ergs}/\mathrm{cm}^2/\mathrm{s}$  \cite{Har-etal:2013ApJ}. Consider a BH with mass $M=10M_\odot$  with a plasma ball with effective temperature $T_{\textrm{eff}}\approx 10^7\mathrm{K}$  and radius $R_0=0.1R_g=0.1GM/c^2\sim1.5\times 10^5\mathrm{cm}$ and corresponding black body luminosity $L\simeq 7\times 10^{25}R_0^2\,\mathrm{ergs}/s$. Let the distance between BH and observer be $D=1\textrm{kpc}$ then, according to formula $F_o=L/(4\pi D^2)$, the value of observer flux will scale as $F_o\simeq 10^{-8}\mathrm{ergs}/\mathrm{cm}^2/\mathrm{s}$. For the echo to be detectable, the ratio between the flux of the primary signal and the echoed signal cannot exceed $10^6$. In our simulation, this ratio is in the range $\sim20-80$. 

While our computation considers photons emitted in all directions, to gain qualitative insight into this phenomenon, let us consider a simple estimation restricted to the equatorial plane, i.e., photons emitted radially inward with zero angular momentum. 
Let us determine the coordinate time $\Delta t_F$ it takes for a complete trip in which a photon travels from the emitter to the mirror and back to the emitter. This estimates the additional delays in receiving the echo signal due to the reflection, which we name the echo delay. Here, the observer is assumed to lie on the same circular orbit as the emitter. The echo delay is given by  
\begin{equation}
    \Delta t_F = 2\int_{u_f}^{u_e}\frac{1+a^2u^2(1+2u)}{u^2(1-2u+a^2u^2)\sqrt{U}}\,du, \label{eq:delay}
\end{equation}
where $u_f=1/r_f=1/(r_h+\epsilon)$ is the reciprocal radius of the mirror, and $u_e=1/r_e$ is the reciprocal radius of the emitter. For comparison, the orbital period of a photon on a circular geodesic at radius $r$ is given by  
\begin{equation}
    \Delta T = \frac{2\pi}{\Omega_e(r)} = 2\pi \left(a + \sqrt{r^3}\right), \label{eq:orbit_period}
\end{equation}
where $\Omega_e(r)$ denotes the orbital angular frequency of the emitter as defined in the expression~\eqref{eq:Omega}. 

In our analysis, we place the hot spot at the innermost stable circular orbit (ISCO), which is also mentioned in the previous section, which provides the most relevant location for strong gravity effects and maximizes the observational impact of potential echoes in the light curve. As seen by a static observer, the trajectory of a photon from ISCO to the mirror surface is shown in Fig.~\ref{fig:echoI}.

The upper panel of Fig.~\ref{fig:echoI} illustrates different scenarios for a representative BH spin of $a=0.9M$ with various mirror depths $\epsilon$, while the lower panel displays 
different scenarios for a representative mirror depth $\epsilon=10^{-6}M$ with various BH spins.
Overall, in all these plots, it can be seen that the photon path gets longer as the distance between the ISCO and the mirror position decreases, which can also be understood in Fig.~\ref{fig:echoII}.

We can see that for larger spin values or lower $\epsilon$ values, the ISCO radius shifts closer to the horizon, causing the photon trajectories to traverse regions of stronger curvature and redshift, with the effect being particularly pronounced for deeper mirrors. 
If a photon takes a longer path to reach the mirror, after the reflection, the photon takes the same-length trajectory to return to the ISCO, which enhances the round-trip travel time and therefore produces longer echo delays.

Since the reflected photons experience a finite time delay, it is not guaranteed that all photons emitted during a given orbital period that are being reflected will be observed within that same period. For instance, photons emitted near the end of an orbital cycle and having suffered a reflection are naturally expected to arrive during subsequent orbits due to the additional travel time. More generally, whether a photon returns within the same orbit or after one or more later orbits depends sensitively on the magnitude of the echo delay.  
If the echo delay is shorter than the orbital period, photons emitted at the beginning of the cycle may still reach the observer within the same orbit, while those emitted later may contribute to subsequent orbits. On the other hand, if the delay exceeds one or more orbital periods, even photons emitted early in the cycle will necessarily be detected only after the hot spot has completed one or several additional revolutions.  
To clarify this behavior, we examine the relation between the echo delay and the orbital timescale, which allows us to determine under what conditions photons reflected from the mirror are observed within the same orbit or instead appear as delayed contributions in later orbital phases.

As shown in the upper-left panel of Fig.~\ref{fig:echoIII}, for $a=0.5M$ and $\epsilon=10^{-5}M$, the echo delay (solid blue line) remains consistently shorter than the orbital period (dashed yellow line). In this case, photons reflected from early hot spot emission are guaranteed to return to the observer within the same orbital period.

In the bottom-left panel ($\epsilon=10^{-1}M$), which considers nearly extremal spin $a=0.998M$, the trend changes with orbital location of the hot spot. 
Close to the ISCO, the echo delay is slightly shorter than the orbital period, but if the radial location of the hot spot moves outward, the echo delay $\Delta t_F$ eventually exceeds $\Delta T$, and at larger radii the orbital period again becomes the dominant timescale. 
This transition highlights the delicate interplay between spin, mirror depth, and orbital location in shaping the timing and structure of the observed echo signal.

A different behavior appears, for example, in the bottom-middle ($\epsilon=0.147M$) and bottom-right ($\epsilon=0.155M$) panels; the orbital period is always larger or comparable to the echo delay. 
This implies that reflected photons from early emission are consistently observed within a single orbital cycle. 
An interesting feature appears in the bottom-middle panel: when the hot spot is located near $\sim 1.5\,r_{\mathrm{ISCO}}$, the echo delay becomes comparable to the orbital period. In this marginal case, photons emitted early in the orbit return to the observer only near the end of the same cycle, producing an overlap between direct and reflected contributions within a single orbital period.

\begin{figure}[H]
    \centering
    \includegraphics[width=0.95\linewidth]{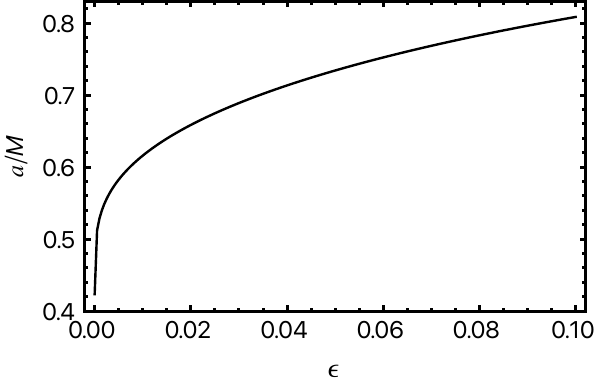}
    \caption{Plot of $a(\epsilon)$ of photon emitted from ISCO, reflecting off the mirror with the value of its depth $\epsilon$ and captured by the hot spot.}
    \label{fig:recapture}
\end{figure}

It is interesting to consider the special case where a zero angular momentum photon emitted from \textcolor{blue}{the} ISCO is reflected by the mirror and captured by the hot spot, i.e. hot spot is at \textcolor{blue}{the} ISCO orbit, where
\begin{equation}
    \Delta \phi=\Omega_k(r_{\mathrm{ISCO}})\Delta t \ ,
\end{equation}
and along photon trajectory $\mathrm{ISCO}\rightarrow\mathrm{mirror}\rightarrow\mathrm{ISCO}$ there is
\begin{eqnarray}
    \Delta t_\mathrm{ISCO}(a,\epsilon)&=&2\int\limits^{u_{ISCO}(a)}_{u_f(a,\epsilon)}\frac{H(u,a)}{\sqrt{U(u,a)}}du \ ,\nonumber \\
    && H(u,a)=\frac{(1+a^2 u^3 (2-a^2 u))}{u^2 (1+u (-2+a^2 u))}\ ,\\
    \Delta\phi_\mathrm{ISCO}(a,\epsilon)&=&2\int\limits^{u_{ISCO}(a)}_{u_f(a,\epsilon)}\frac{P(u,a)}{\sqrt{U(u,a)}}du \ ,\label{phi_gamma} \nonumber \\ && P(u,a)=\frac{2 a u}{1-2 u+a^2 u^2} \ ,
\end{eqnarray}
with
\begin{equation}
    U(u,a)=1+a^2u^2(1+2u).
\end{equation}
We are looking for such $(a,\epsilon)$ satisfying the equation
\begin{equation}
    \Delta t_\mathrm{ISCO}(a,\epsilon)\Omega_k(r_\mathrm{ISCO}(a))-\Delta\phi_\mathrm{ISCO}(a,\epsilon)=0.
\end{equation}
It results in the construction of a function of Kerr spin $a=a(\epsilon)$ for this effect, resulting in the plot we show in Fig.~\ref{fig:recapture}.
For a given mirror depth $\epsilon$, there is a value of Kerr spin $a$ such that a zero angular momentum photon will not contribute to the reflected image. In analogy, one can find such a map for any photon. 

The sensitivity of our results to the integration scheme and tolerances is studied. For Eq.~\eqref{phi_gamma}, comparison against an arbitrary-precision~(40-digit) quadrature and four independent solvers~(explicit Runge–Kutta of orders five and eight, implicit Runge–Kutta, variable-order backward differentiation) gives agreement to a relative $\sim 10^ {-10}$ or better.
It is important to emphasize that these discussions are based on a simplified treatment, considering only radial photons with zero angular momentum. 
In more realistic scenarios, photons with non-zero angular momentum and non-radial trajectories will exhibit more complex delay structures, potentially modifying the detailed shape and timing of the echo signal. 
Nevertheless, the qualitative picture remains robust: the interplay between the orbital period of the hot spot and the round-trip photon delay to the mirror governs whether echoes reinforce radiation within the same orbit or contribute to subsequent orbital cycles.

\section{Discussion and Conclusion} \label{discussion}

In the case of a Kerr BH, some fraction of photons that are emitted from the hot spot may be lost down to the horizon of the BH forever. If the event horizon is completely screened by the mirror, then the emitted photon would suffer reflection when it hits the mirror. A fraction of these photons would travel up to the observer and contribute to the observed flux.

We perform ray-tracing simulations to construct the geodesic of the photons which travel up to the observer from the hot spot, and considering the local Gaussian-like emission profile as given in~\eqref{eq:I}. In the simulations, to get the maximum effect of the reflection, we stick to the orbit of the hot spot on the equatorial plane at $r=r_{_{\mathrm{ISCO}}}$, for an observer's inclination angle, $\theta_o=30\degree$, and the reflective surface is put close to the horizon, i.e., $\epsilon=10^{-6}M$.
While the hot spot remains on the equatorial plane, the photon emitted from it is assumed to be isotropic and treated accordingly in the simulations.

In Fig.~\ref{fig:spineffect}, one can see that the echo slightly enhances the observed light curve. This is a scenario when the hot spot ceases to exist after three revolutions. An echo tail due to reflected photons is observed at the end of the primary emission. The reason for this enhancement is that the effective potential well gets deeper for a higher value of spin, and the photon suffers a longer echo delay. Also, as the spin parameter increases, the orbit, i.e., the ISCO, gets closer and closer to the horizon, thus the mirror. So, the emitted photons are much closer to the deep potential well of the Kerr mirror.

We demonstrate how the depth of the mirror affects the shape of the light curve in Fig.~\ref{fig:maxspindiffeps} for a rapidly rotating BH with maximal spin $a=0.998M$. In this figure, we show the light curve for six representative values of the depths of the mirror, $\epsilon=10^{-1}M$, $10^{-2}M$, $10^{-3}M$, $10^{-4}M$, $10^{-5}M$, and $10^{-6}M$. For a fixed BH spin, if we put the mirror closer to the horizon, a longer echo is seen, as expected. After the hot spot ceases to exist, an echo wave-train can be observed.

Since the flux enhancement produced by the echo is relatively modest, its imprint on the light curve may be difficult to disentangle when the hot spot remains on a stable orbit, like a star in a stable orbit around a supermassive BH for an extended period. In contrast, the effect could become more pronounced for transient hot spots that form and dissipate within the accretion flow on short timescales. Another potentially interesting scenario arises when the hot spot fails to maintain a stable orbit and plunges into the mirror, like a scenario in which a star is spiraling down to a supermassive BH, leading to a collision with the mirror. Investigating such dynamics is beyond the scope of the present work and is left for future study.

We see that the reflected photon contribution in the observed light curve appears at a late time, very similar to gravitational wave echoes \cite{Ferrari2000,Abedi2017,Cardoso2017}. Reflected photons take very different times to reach the observer for different configurations of the system, as extensively discussed in the previous section. The echo delay lengthens both as the mirror is placed closer to the horizon and as the spin is increased, and once the hot spot ceases to exist, the reflected contribution is left as an echo wave-train at the end of the primary emission.

We stress that a sophisticated quantitative study considering the hot spot in a general orbit is necessary to investigate a more realistic astrophysical scenario and whether such a signature could be observed with current facilities. A more detailed study in this direction will be conducted in future work. This analysis, considering a toy model, has the merit of indicating new features that appear due to the presence of a mirror in the light curve.

\section{Acknowledgement}
S.H., J.S., and Z.S. thank the Institute of Physics, Silesian University in Opava, for the institutional support provided during this work.
This work was supported by the Student Grant Foundation of the Silesian University in Opava, Grant No. SGF/2/2021, which was realized within the EU OPSRE project entitled "Improving the quality of the internal grant scheme of the Silesian University in Opava", reg. number: \\CZ.02.2.69/0.0/0.0/19\_073/0016951. The authors would like to thank Enrico Barausse for his valuable comments during the initial phase of the project.
We thank the anonymous referee for the constructive comments and suggestions, which helped improve the clarity of the manuscript.

\bibliographystyle{unsrt}
\bibliography{ref}

@article{2016,
   title={Observation of Gravitational Waves from a Binary Black Hole Merger},
   volume={116},
   ISSN={1079-7114},
   url={http://dx.doi.org/10.1103/PhysRevLett.116.061102},
   DOI={10.1103/physrevlett.116.061102},
   number={6},
   journal={Physical Review Letters},
   publisher={American Physical Society (APS)},
   author={Abbott et. al},
   year={2016},
   month={Feb} }

@article{EventHorizonTelescope:2019dse,
    author = "Akiyama, Kazunori and others",
    collaboration = "Event Horizon Telescope",
    title = "{First M87 Event Horizon Telescope Results. I. The Shadow of the Supermassive Black Hole}",
    eprint = "1906.11238",
    archivePrefix = "arXiv",
    primaryClass = "astro-ph.GA",
    doi = "10.3847/2041-8213/ab0ec7",
    journal = "Astrophys. J. Lett.",
    volume = "875",
    pages = "L1",
    year = "2019"
}

@article{EventHorizonTelescope:2019ths,
    author = "Akiyama, Kazunori and others",
    collaboration = "Event Horizon Telescope",
    title = "{First M87 Event Horizon Telescope Results. IV. Imaging the Central Supermassive Black Hole}",
    eprint = "1906.11241",
    archivePrefix = "arXiv",
    primaryClass = "astro-ph.GA",
    doi = "10.3847/2041-8213/ab0e85",
    journal = "Astrophys. J. Lett.",
    volume = "875",
    number = "1",
    pages = "L4",
    year = "2019"
}

@article{EventHorizonTelescope:2021bee,
    author = "Akiyama, Kazunori and others",
    collaboration = "Event Horizon Telescope",
    title = "{First M87 Event Horizon Telescope Results. VII. Polarization of the Ring}",
    eprint = "2105.01169",
    archivePrefix = "arXiv",
    primaryClass = "astro-ph.HE",
    doi = "10.3847/2041-8213/abe71d",
    journal = "Astrophys. J. Lett.",
    volume = "910",
    number = "1",
    pages = "L12",
    year = "2021"
}

@article{EventHorizonTelescope:2021srq,
    author = "Akiyama, Kazunori and others",
    collaboration = "Event Horizon Telescope",
    title = "{First M87 Event Horizon Telescope Results. VIII. Magnetic Field Structure near The Event Horizon}",
    eprint = "2105.01173",
    archivePrefix = "arXiv",
    primaryClass = "astro-ph.HE",
    doi = "10.3847/2041-8213/abe4de",
    journal = "Astrophys. J. Lett.",
    volume = "910",
    number = "1",
    pages = "L13",
    year = "2021"
}

@article{Hawking:1975vcx,
    author = "Hawking, S. W.",
    editor = "Gibbons, G. W. and Hawking, S. W.",
    title = "{Particle Creation by Black Holes}",
    doi = "10.1007/BF02345020",
    journal = "Commun. Math. Phys.",
    volume = "43",
    pages = "199--220",
    year = "1975",
    note = "[Erratum: Commun.Math.Phys. 46, 206 (1976)]"
}

@article{Almheiri:2012rt,
    author = "Almheiri, Ahmed and Marolf, Donald and Polchinski, Joseph and Sully, James",
    title = "{Black Holes: Complementarity or Firewalls?}",
    eprint = "1207.3123",
    archivePrefix = "arXiv",
    primaryClass = "hep-th",
    doi = "10.1007/JHEP02(2013)062",
    journal = "JHEP",
    volume = "02",
    pages = "062",
    year = "2013"
}

@ARTICLE{1992A&A...257..594B,
       author = {{Bao}, G.},
        title = "{The signature of corotating spots in accretion disks}",
      journal = {Astronomy and Astrophysics},
         year = 1992,
        month = apr,
       volume = {257},
       number = {2},
        pages = {594-598},
       adsurl = {https://ui.adsabs.harvard.edu/abs/1992A&A...257..594B}
}

@article{Stuchlik:1992phc,
    author = "Stuchlik, Z. and Bao, G.",
    title = "{Radiation from hot spots orbiting an extreme Reissner-Nordstrom black hole}",
    reportNumber = "SISSA-66-92-A",
    journal = "Gen. Rel. Grav.",
    volume = "24",
    pages = "945--957",
    year = "1992"
}

@article{ refId0,
	author = {{GRAVITY Collaboration: Abuter, R., Amorim, A., et al.}},
	title = {Detection of orbital motions near the last stable circular orbit of the massive black hole SgrA*},
	DOI= "10.1051/0004-6361/201834294",
	url= "https://doi.org/10.1051/0004-6361/201834294",
	journal = {A\&A},
	year = 2018,
	volume = 618,
	pages = "L10",
}

@article{10.1093/mnras/269.2.283,
    author = {Zakharov, A. F.},
    title = "{On the hotspot near a Kerr black hole: Monte-Carlo simulations}",
    journal = {Monthly Notices of the Royal Astronomical Society},
    volume = {269},
    number = {2},
    pages = {283-288},
    year = {1994},
    month = {07},
    issn = {0035-8711},
    doi = {10.1093/mnras/269.2.283},
    url = {https://doi.org/10.1093/mnras/269.2.283},
    eprint = {https://academic.oup.com/mnras/article-pdf/269/2/283/3150595/mnras269-0283.pdf},
}

@INPROCEEDINGS{2004AIPC..714...40S,
       author = {{Schnittman}, Jeremy and {Bertschinger}, Edmund},
        title = "{A Hot Spot Model for Black Hole QPOs}",
    booktitle = {X-ray Timing 2003: Rossi and Beyond},
         year = 2004,
       editor = {{Kaaret}, Philip and {Lamb}, Frederick K. and {Swank}, Jean H.},
       series = {American Institute of Physics Conference Series},
       volume = {714},
        month = jul,
        pages = {40-43},
          doi = {10.1063/1.1780996},
archivePrefix = {arXiv},
       eprint = {astro-ph/0312406},
 primaryClass = {astro-ph},
       adsurl = {https://ui.adsabs.harvard.edu/abs/2004AIPC..714...40S}
}

@article{Li:2014coa,
    author = "Li, Zilong and Bambi, Cosimo",
    title = "{Distinguishing black holes and wormholes with orbiting hot spots}",
    eprint = "1405.1883",
    archivePrefix = "arXiv",
    primaryClass = "gr-qc",
    doi = "10.1103/PhysRevD.90.0242000071",
    journal = "Phys. Rev. D",
    volume = "90",
    pages = "024071",
    year = "2014"
}

@ARTICLE{2000MNRAS.315....1R,
       author = {{Ruszkowski}, Mateusz},
        title = "{X-ray iron line variability for the model of an orbiting flare above a black hole accretion disc}",
      journal = {Monthly Notices of the Royal Astronomical Society},
         year = 2000,
        month = jun,
       volume = {315},
       number = {1},
        pages = {1-10},
          doi = {10.1046/j.1365-8711.2000.02898.x},
archivePrefix = {arXiv},
       eprint = {astro-ph/9906397},
 primaryClass = {astro-ph},
       adsurl = {https://ui.adsabs.harvard.edu/abs/2000MNRAS.315....1R}
}

@ARTICLE{1996ApJ...470..743K,
       author = {{Karas}, V.},
        title = "{Light Curve of a Source Orbiting a Black Hole: A Fitting Formula}",
      journal = {Astrophysical Journal},
         year = 1996,
        month = oct,
       volume = {470},
        pages = {743},
          doi = {10.1086/177904},
archivePrefix = {arXiv},
       eprint = {astro-ph/9605083},
 primaryClass = {astro-ph},
       adsurl = {https://ui.adsabs.harvard.edu/abs/1996ApJ...470..743K}
}

@INPROCEEDINGS{1995NYASA.759..550Z,
       author = {{Zakharov}, Alexander},
        title = "{On the Hot Spot near a Kerr Black Hole.}",
    booktitle = {Seventeeth Texas Symposium on Relativistic Astrophysics and Cosmology},
         year = 1995,
       editor = {{B{\"o}hringer}, Hans and {Morfill}, Gregor E. and {Tr{\"u}mper}, Joachim E.},
       volume = {759},
        month = jan,
        pages = {550},
          doi = {10.1111/j.1749-6632.1995.tb17606.x},
       adsurl = {https://ui.adsabs.harvard.edu/abs/1995NYASA.759..550Z}
}

@ARTICLE{1973ApJ...183..237C,
       author = {{Cunningham}, C.~T. and {Bardeen}, James M.},
        title = "{The Optical Appearance of a Star Orbiting an Extreme Kerr Black Hole}",
      journal = {Astrophysical Journal},
         year = 1973,
        month = jul,
       volume = {183},
        pages = {237-264},
          doi = {10.1086/152223},
       adsurl = {https://ui.adsabs.harvard.edu/abs/1973ApJ...183..237C}
}

@ARTICLE{2006AN....327..957P,
       author = {{Pech{\'a}{\v{c}}ek}, T. and {Dov{\v{c}}iak}, M. and {Karas}, V.},
        title = "{Power spectra from `spotted' accretion discs}",
      journal = {Astronomische Nachrichten},
         year = 2006,
        month = dec,
       volume = {327},
       number = {10},
        pages = {957},
          doi = {10.1002/asna.200610671},
       adsurl = {https://ui.adsabs.harvard.edu/abs/2006AN....327..957P}
}

@ARTICLE{2003PASJ...55.1121F,
       author = {{Fukue}, Jun},
        title = "{Light-Curve Diagnosis of a Hot Spot for Accretion-Disk Models}",
      journal = {Publications of the Astronomical Society of Japan},
         year = 2003,
        month = dec,
       volume = {55},
        pages = {1121-1125},
          doi = {10.1093/pasj/55.6.1121},
       adsurl = {https://ui.adsabs.harvard.edu/abs/2003PASJ...55.1121F}
}

@ARTICLE{1994ApJ...425...63B,
       author = {{Bao}, G. and {Hadrava}, P. and {Ostgaard}, E.},
        title = "{Multiple Images and Light Curves of an Emitting Source on a Relativistic Eccentric Orbit around a Black Hole}",
      journal = {Astrophysical Journal},
         year = 1994,
        month = apr,
       volume = {425},
        pages = {63},
          doi = {10.1086/173963},
       adsurl = {https://ui.adsabs.harvard.edu/abs/1994ApJ...425...63B}
}

@article{Cardoso2017,
    author = "Cardoso, Vitor and Pani, Paolo",
    title = "{Tests for the existence of black holes through gravitational wave echoes}",
    eprint = "1709.01525",
    archivePrefix = "arXiv",
    primaryClass = "gr-qc",
    doi = "10.1038/s41550-017-0225-y",
    journal = "Nature Astron.",
    volume = "1",
    number = "9",
    pages = "586--591",
    year = "2017"
}

@article{Ferrari2000,
  title = {Scattering of particles by neutron stars: Time evolutions for axial perturbations},
  author = {Ferrari, V. and Kokkotas, K. D.},
  journal = {Phys. Rev. D},
  volume = {62},
  issue = {10},
  pages = {107504},
  numpages = {4},
  year = {2000},
  month = {Oct},
  publisher = {American Physical Society},
  doi = {10.1103/PhysRevD.62.107504},
  url = {https://link.aps.org/doi/10.1103/PhysRevD.62.107504}
}

@article{Abedi2017,
    author = "Abedi, Jahed and Dykaar, Hannah and Afshordi, Niayesh",
    title = "{Echoes from the Abyss: Tentative evidence for Planck-scale structure at black hole horizons}",
    eprint = "1612.00266",
    archivePrefix = "arXiv",
    primaryClass = "gr-qc",
    doi = "10.1103/PhysRevD.96.082004",
    journal = "Phys. Rev. D",
    volume = "96",
    number = "8",
    pages = "082004",
    year = "2017"
}

@article{Carter1968,
    author = "Carter, Brandon",
    title = "{Hamilton-Jacobi and Schrodinger Separable Solutions of Einstein's Equations}",
    doi = "10.1007/BF03399503",
    journal = "Commu. in Math. Phys.",
    volume = "10",
    number = "4",
    pages = "pp. 280-310",
    year = "1968"
}

@article{Bardeen1972,
    author = "Bardeen, James M. and  Press, William H. and  Teukolsky, Saul A.",
    title = "{Rotating Black Holes: Locally Nonrotating Frames, Energy Extraction, and Scalar Synchrotron Radiation}",
    doi = "10.1086/151796",
    journal = "Astrophys. J.",
    volume = "178",
    pages = "pp. 347-370",
    year = "1972"
}

@article{Bao-Stu:1992:ApJ,
    author = "Bao, G. and  Stuchlik, Z.",
    title = "Accretion Disk Self-Eclipse: X-Ray Light Curve and Emission Line",
    journal = "Astrophys. J.",
    volume = "400",
    pages = "p.163",
    year = "1992"
}

@article{Mazur:2001fv,
    author = "Mazur, Pawel O. and Mottola, Emil",
    title = "{Gravitational Condensate Stars: An Alternative to Black Holes}",
    eprint = "gr-qc/0109035",
    archivePrefix = "arXiv",
    reportNumber = "LA-UR-01-5067, LA-UR-01-5067",
    doi = "10.3390/universe9020088",
    journal = "Universe",
    volume = "9",
    number = "2",
    pages = "88",
    year = "2023"
}

@article{Visser:2003ge,
    author = "Visser, Matt and Wiltshire, David L.",
    title = "{Stable gravastars: An Alternative to black holes?}",
    eprint = "gr-qc/0310107",
    archivePrefix = "arXiv",
    doi = "10.1088/0264-9381/21/4/027",
    journal = "Class. Quant. Grav.",
    volume = "21",
    pages = "1135--1152",
    year = "2004"
}

@article{Liebling:2012fv,
    author = "Liebling, Steven L. and Palenzuela, Carlos",
    title = "{Dynamical boson stars}",
    eprint = "1202.5809",
    archivePrefix = "arXiv",
    primaryClass = "gr-qc",
    doi = "10.1007/s41114-023-00043-4",
    journal = "Living Rev. Rel.",
    volume = "26",
    number = "1",
    pages = "1",
    year = "2023"
}

@article{Cardoso:2019rvt,
    author = "Cardoso, Vitor and Pani, Paolo",
    title = "{Testing the nature of dark compact objects: a status report}",
    eprint = "1904.05363",
    archivePrefix = "arXiv",
    primaryClass = "gr-qc",
    doi = "10.1007/s41114-019-0020-4",
    journal = "Living Rev. Rel.",
    volume = "22",
    number = "1",
    pages = "4",
    year = "2019"
}

@article{Punturo:2010zz,
    author = "Punturo, M. and others",
    editor = "Ricci, Fulvio",
    title = "{The Einstein Telescope: A third-generation gravitational wave observatory}",
    doi = "10.1088/0264-9381/27/19/194002",
    journal = "Class. Quant. Grav.",
    volume = "27",
    pages = "194002",
    year = "2010"
}

@article{Reitze:2019iox,
    author = "Reitze, David and others",
    title = "{Cosmic Explorer: The U.S. Contribution to Gravitational-Wave Astronomy beyond LIGO}",
    eprint = "1907.04833",
    archivePrefix = "arXiv",
    primaryClass = "astro-ph.IM",
    reportNumber = "LIGO-P1900316",
    journal = "Bull. Am. Astron. Soc.",
    volume = "51",
    number = "7",
    pages = "035",
    year = "2019"
}

@ARTICLE{2017arXiv170200786A,
       author = {{Amaro-Seoane}, Pau and {Audley}, Heather and {Babak}, Stanislav and {Baker}, John and {Barausse}, Enrico and {Bender}, Peter },
        title = "{Laser Interferometer Space Antenna}",
      journal = {arXiv e-prints},
         year = 2017,
        month = feb,
          eid = {arXiv:1702.00786},
        pages = {arXiv:1702.00786},
          doi = {10.48550/arXiv.1702.00786},
archivePrefix = {arXiv},
       eprint = {1702.00786},
 primaryClass = {astro-ph.IM},
       adsurl = {https://ui.adsabs.harvard.edu/abs/2017arXiv170200786A}
}

@INPROCEEDINGS{2019BAAS...51g.256D,
       author = {{Doeleman}, Sheperd and {Blackburn}, Lindy and {Dexter}, Jason and {Gomez}, Jose L. and {Johnson}},
        title = "{Studying Black Holes on Horizon Scales with VLBI Ground Arrays}",
    booktitle = {Bulletin of the American Astronomical Society},
         year = 2019,
       volume = {51},
        month = sep,
          eid = {256},
        pages = {256},
          doi = {10.48550/arXiv.1909.01411},
archivePrefix = {arXiv},
       eprint = {1909.01411},
 primaryClass = {astro-ph.IM},
       adsurl = {https://ui.adsabs.harvard.edu/abs/2019BAAS...51g.256D}
}

@article{Schnittman:2003tm,
    author = "Schnittman, Jeremy D. and Bertschinger, Edmund",
    title = "{A model for high frequency quasi-periodic oscillations from accreting black holes}",
    eprint = "astro-ph/0309458",
    archivePrefix = "arXiv",
    doi = "10.1086/383180",
    journal = "Astron. Astrophys.",
    volume = "606",
    pages = "1098",
    year = "2004"
}

@article{Stu-Vrb:2026:PhRvD,
    author = "Stuchl\'{i}k, Zden\v{e}k and Vrba, Jaroslav",
    title = "{Magnetosphere of black hole mimickers: Charged particle belts and repulsive barrier inside the photon sphere}",
    doi = "10.1103/fwdw-gpyz",
    journal = "Phys. Rev. D",
    volume = "113, 12, id.123013",
    pages = "22",
    year = "2026"
}

@article{Stu-Vrb-Tur:2026:EPJC,
    author = "Stuchl\'{i}k, Zden\v{e}k and Vrba, Jaroslav and Tursunov, Arman",
    title = "{Charged particles orbiting black hole mimickers with dipole magnetosphere }",
    doi = "10.1140/epjc/s10052-026-15731-y",
    journal = "The Europ. Phys. Jour. C",
    volume = "86, 5, id.562",
    year = "2026"
}

@ARTICLE{2012A&A...546A.121P,
       author = {{Potekhin}, A.~Y. and {Suleimanov}, V.~F. and {van Adelsberg}, M. and {Werner}, K.},
        title = "{Radiative properties of magnetic neutron stars with metallic surfaces and thin atmospheres}",
      journal = "Astronomy \& Astrophysics",
         year = 2012,
        month = oct,
       volume = {546},
          eid = {A121},
        pages = {A121},
          doi = {10.1051/0004-6361/201219747},
archivePrefix = {arXiv},
       eprint = {1208.6582},
 primaryClass = {astro-ph.HE},
       adsurl = {https://ui.adsabs.harvard.edu/abs/2012A&A...546A.121P}
}

@ARTICLE{Fan-etal:1997PASJ,
       author = {{Fanton}, Claudio and {Calvani}, Massimo and {de Felice}, Fernando and {Cadez}, Andrej},
        title = "{Detecting Accretion Disks in Active Galactic Nuclei}",
      journal = "Publications of the Astronomical Society of Japan",
         year = 1997,
        month = apr,
       volume = {49},
        pages = {159-169},
          doi = {10.1093/pasj/49.2.159},
       adsurl = {https://ui.adsabs.harvard.edu/abs/1997PASJ...49..159F}
}

@ARTICLE{Schee-Stu:2009GRG,
       author = {{Schee}, Jan and {Stuchl{\'\i}k}, Zden{\v{e}}k},
        title = "{Profiles of emission lines generated by rings orbiting braneworld Kerr black holes}",
      journal = {General Relativity and Gravitation},
         year = 2009,
        month = aug,
       volume = {41},
       number = {8},
        pages = {1795-1818},
          doi = {10.1007/s10714-008-0753-y},
archivePrefix = {arXiv},
       eprint = {0812.3017},
 primaryClass = {astro-ph},
       adsurl = {https://ui.adsabs.harvard.edu/abs/2009GReGr..41.1795S}
}

@ARTICLE{Har-etal:2013ApJ,
       author = {{Harrison}, Fiona A. and {Craig}, William W. and {Christensen}, Finn E. and {Hailey}, Charles J. and {Zhang}, William W. and {Boggs}, Steven E. and {Stern}, Daniel and {Cook}, W. Rick and {Forster}, Karl and {Giommi}, Paolo and {Grefenstette}, Brian W. and {Kim}, Yunjin and {Kitaguchi}, Takao and {Koglin}, Jason E. and {Madsen}, Kristin K. and {Mao}, Peter H. and {Miyasaka}, Hiromasa and {Mori}, Kaya and {Perri}, Matteo and {Pivovaroff}, Michael J. and {Puccetti}, Simonetta and {Rana}, Vikram R. and {Westergaard}, Niels J. and {Willis}, Jason and {Zoglauer}, Andreas and {An}, Hongjun and {Bachetti}, Matteo and {Barri{\`e}re}, Nicolas M. and {Bellm}, Eric C. and {Bhalerao}, Varun and {Brejnholt}, Nicolai F. and {Fuerst}, Felix and {Liebe}, Carl C. and {Markwardt}, Craig B. and {Nynka}, Melania and {Vogel}, Julia K. and {Walton}, Dominic J. and {Wik}, Daniel R. and {Alexander}, David M. and {Cominsky}, Lynn R. and {Hornschemeier}, Ann E. and {Hornstrup}, Allan and {Kaspi}, Victoria M. and {Madejski}, Greg M. and {Matt}, Giorgio and {Molendi}, Silvano and {Smith}, David M. and {Tomsick}, John A. and {Ajello}, Marco and {Ballantyne}, David R. and {Balokovi{\'c}}, Mislav and {Barret}, Didier and {Bauer}, Franz E. and {Blandford}, Roger D. and {Brandt}, W. Niel and {Brenneman}, Laura W. and {Chiang}, James and {Chakrabarty}, Deepto and {Chenevez}, Jerome and {Comastri}, Andrea and {Dufour}, Francois and {Elvis}, Martin and {Fabian}, Andrew C. and {Farrah}, Duncan and {Fryer}, Chris L. and {Gotthelf}, Eric V. and {Grindlay}, Jonathan E. and {Helfand}, David J. and {Krivonos}, Roman and {Meier}, David L. and {Miller}, Jon M. and {Natalucci}, Lorenzo and {Ogle}, Patrick and {Ofek}, Eran O. and {Ptak}, Andrew and {Reynolds}, Stephen P. and {Rigby}, Jane R. and {Tagliaferri}, Gianpiero and {Thorsett}, Stephen E. and {Treister}, Ezequiel and {Urry}, C. Megan},
        title = "{The Nuclear Spectroscopic Telescope Array (NuSTAR) High-energy X-Ray Mission}",
      journal = {\apj},
         year = 2013,
        month = jun,
       volume = {770},
       number = {2},
          eid = {103},
        pages = {103},
          doi = {10.1088/0004-637X/770/2/103},
archivePrefix = {arXiv},
       eprint = {1301.7307},
 primaryClass = {astro-ph.IM},
       adsurl = {https://ui.adsabs.harvard.edu/abs/2013ApJ...770..103H}
}

\end{document}